\documentclass[utf8]{frontiersFPHY} 
\usepackage{url,hyperref,lineno,microtype,subcaption}
\usepackage[onehalfspacing]{setspace}

\usepackage{braket}

\def\keyFont{\fontsize{8}{11}\helveticabold }
\def\firstAuthorLast{Hern\'{a}ndez-G\'{omez} and Fabbri} 
\def\Authors{Santiago Hern\'{a}ndez-G\'{omez}\,$^{1}$, and Nicole Fabbri\,$^{1,*}$}
\def\Address{$^{1}$LENS European Laboratory for Non-linear Spectroscopy, and CNR-INO Istituto Nazionale di Ottica, Sesto Fiorentino, Italy}
\def\corrAuthor{Nicole Fabbri, via N. Carrara 1, I-50019 Sesto Fiorentino (FI), Italy} 
\def\corrEmail{fabbri@lens.unifi.it}

\begin{document}
\onecolumn
\firstpage{1}

\title[Nanoscale spectroscopy with diamond NV-centers]{Quantum control for nanoscale spectroscopy with diamond NV centers: A short review}
\author[\firstAuthorLast ]{\Authors} 
\address{} 
\correspondance{} 

\extraAuth{}
\maketitle
\begin{abstract}
Diamond quantum technologies based on color centers have rapidly emerged in the most recent years. The Nitrogen-Vacancy (NV) color center has  attracted a particular interest thanks to its outstanding spin properties and optical addressability. The NV center has been used to realize innovative multi-mode quantum-enhanced sensors that offer an unprecedented combination of high sensitivity and spatial resolution at room temperature. The technological progress and the widening of potential sensing applications has induced an increasing demand for performance advances of NV quantum sensors. Quantum control (QC) plays a key role in responding to this demand. This short review affords an overview on recent advances in QC-assisted quantum sensing and spectroscopy of magnetic fields.

\tiny
 \keyFont{ \section{Keywords:} NV centers in diamond, quantum devices, quantum sensors, quantum control, noise spectroscopy}
\end{abstract}

\section{Introduction}
Optically-active point defects in diamond, so-called color centers, have drawn a general interest in the field of quantum technologies the last few years, thanks to their attractive and variate capabilities
~\cite{Bradac20,Thiering20}. 
Among diamond color centers, the negatively charged Nitrogen-Vacancy (NV) center~\cite{Dobrovitski13, Doherty13} has stood out as a solid-state spin qubit thanks to a high degree of coherent control, ultra-long spin coherence time, remarkable fluorescence photostability, as well as optical addressability, initialization and readout, all of which can be achieved at room temperature. The wide range of applications for the NV center includes its use for quantum memories as building blocks of solid-state quantum registers~\cite{Neumann08,Bradley19}, biocompatible quantum sensors~\cite{Rondin14}, reliable nonclassical sources of single photons~\cite{Sipahigil12}. Very recently NV centers have been also employed as a platform to tackle novel challenges in the investigation of quantum thermodynamics for open systems~\cite{Klatzow19,HernandezGomez20}. 

As quantum sensors, NV-based sensors exploit quantum resources to enhance the detection of physical signals. They have been successfully used for measuring magnetic and electric fields~\cite{Balasubramanian09,Wolf15,Dolde11}, temperature\cite{Acosta10,Neumann13}, rotation~\cite{soshenko2020nuclear}, strain and pressure~\cite{Teissier14}, and more. Remarkably, NV magnetometers have been demonstrated to be capable of measuring very localized ultra weak AC fields, achieving sensitivities of the order of pT/$\sqrt{\mathrm{Hz}}$~\cite{Acosta10}  in ambient conditions. 

However, as for any other practical device, the operation of NV-based quantum sensor is prone to limitations and imperfections: The high sensitivity to the magnetic environment make them very precise sensors, but the same interaction with the environment constitutes also a limit to the device sensitivity by reducing the coherence of the quantum states. Along with progress in the diamond fabrication~\cite{Markham11,Waldermann07}, recent advances in Quantum Control (QC) offers powerful tools to overcome these limitations. This review paper provides a swift compendium of  current advances in QC methods and describes how they improve the NV-sensor sensitivity. The content is organized as follows: Sec.~\ref{magnetometrypinciples} describes the background of NV-magnetometry, 
with Sec.~\ref{properties} introducing the most relevant properties of diamond NV centers, and Sec.~\ref{magn} describing basic DC- and AC-magnetometry schemes.
Sec.~\ref{magn} offers an overview on the design of QC protocols for NV-center magnetometry and their main applications. Throughout the review, we refer to quantum sensing and spectroscopy schemes that exploit state superpositions as a quantum resource --- as opposed to quantum metrology that uses entangled states~\cite{Degen17}. Albeit these schemes have been so far mostly employed with single-spin sensing-qubits in bulk diamond~\cite{DeLange10,Bar-Gill13}, they have also emerged as effective tools for nanodiamond-based~\cite{Knowles14} and ensemble devices~\cite{Pham12,Bar-Gill12}.
 
\section{NV magnetometry}\label{magnetometrypinciples}

\subsection{The NV center in diamond}\label{properties}
The NV center is formed by a substitutional nitrogen atom adjacent to a vacancy in the diamond lattice, with $C_{3v}$ symmetry around one of the four $[111]$ crystallographic directions. In the negatively charged NV$^-$---hereafter referred as NV for simplicity, the favorable internal energy structure and photophysics~\cite{Doherty13} enable optical initialization and readout, and coherent manipulation with long coherence time, opening the way for many quantum technology applications. 

\begin{figure}[h!]
\begin{center}
\includegraphics[width=.85\textwidth]{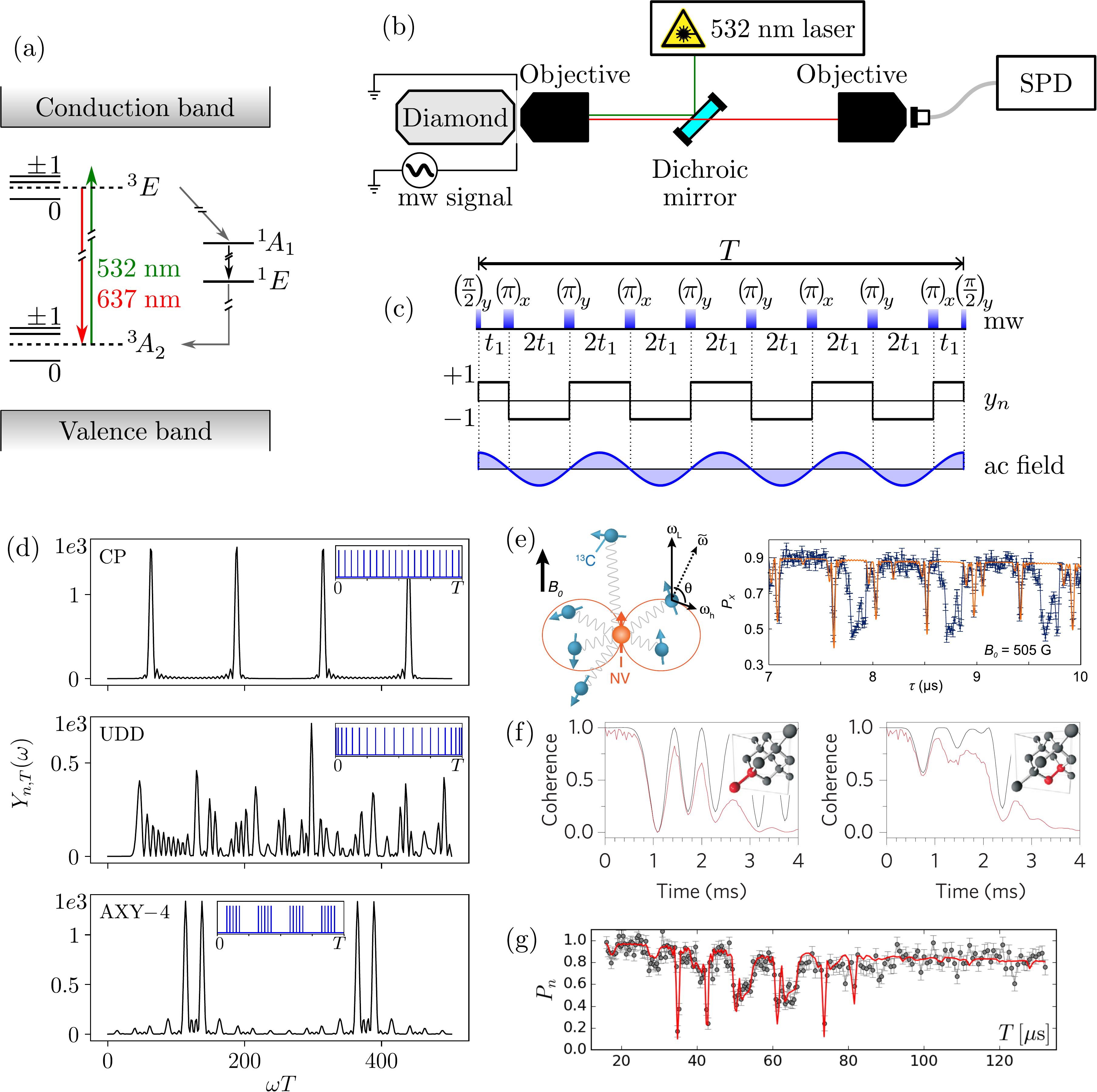}
\end{center}
\caption{\textbf{(a)} Energy level structure of the NV center. 
\textbf{(b)} Simplified scheme of a typical NV experiment implementation. The NV center is addressed with a green laser light via a confocal microscope. The fluorescence emitted by the color center is collected by a single photon detector. An antenna placed in proximity of the diamond chip delivers resonant microwave radiation to control the NV spin dynamics. 
\textbf{(c)} Microwave control pulses distribution of an XY-8 sequence. The qubit acquired phase $\varphi_n(t)$, is maximized when the filter function $y_n$ describing the control protocol is synchronized with the target AC field $b(t)$. \textbf{(d)} Filter function of CP, UDD and  AXY-4 control protocols (see text). Inset: temporal distribution of the $\pi$ pulses. For simplicity, the number of $\pi$ pulses is fixed to $n=20$. 
\textbf{(e-g)} NV spin dynamics under DD sequences exemplified in (d).
\textbf{(e)} Illustrative scheme of the NV center electronic spin, and six nearby $^{13}$C nuclear spins. The precession of the nuclear spins depends on their orientation with respect to the NV axis. The right panel shows experimental data relative to the residual population of the $\ket{1}$ state for a DD protocol using a CP-based sequence with $n=32$ equidistant pulses, and total sensing time is $T=32 (2t_1)$. The orange line represents the prediction after characterizing the interaction with six individual nearby nuclear spins. Adapted from Ref.~\cite{Taminiau12}.
\textbf{(f)} NV spin coherence under a UDD sequence ($n=5$) in the presence of a $^{13}$C dimer, placed $1.1$~nm away from the NV spin (red line), and NV spin coherence once subtracted the effect of the noise-induced decoherence (black line). Each panel refers to a different orientation of the dimer (see inset), with respect to the NV axis. Adapted from Ref.~\cite{Zhao11}. 
\textbf{(g)} Residual population of the $\ket{1}$ state under an AXY-8 sequence ($n=40$). The red line shows the predicted dynamics for the interaction with three resolved nuclear spins and with the spin-bath. Adapted from Ref.~\cite{HernandezGomez18}
}
\label{fig}
\end{figure}

The NV energy structure, shown in Fig.~\ref{fig}(a), consists in electronic orbital ground ($^3A_2$) and excited ($^3E$) triplet levels separated by 1.945 eV,  and two intermediate $^1E$ and $^1A_1$ singlet levels~\cite{Doherty13}. 
Within the spin-1 triplet ground state, the spin projection $m_s = 0$ is separated from  the degenerate $m_S=\pm1$ owing to electronic spin-spin interaction within the NV, with zero-field-splitting of $D_g\simeq2.87$~GHz. A static bias magnetic field further splits the levels $m_S=\pm1$ and modifies their energies via Zeeman effect. A microwave excitation can be used to selectively address one of the $m_S=0\rightarrow\pm1$ transition. Thus, the NV center at room temperature can be effectively employed as a single-qubit probe system, with $|0\rangle$ and $|1\rangle$ being a pair of spin projections. The success of this platform is primarily due to its remarkably long spin coherence-time compared to any other solid-state platform. In most of NV implementations, dephasing is induced by the slowly-varying inhomogeneity of the dipolar fields due to unpolarized spin impurities ($^{13}$C and $^{14}$N) within the diamond crystal. In type II-a CVD-grown bulk diamond with Carbon natural abundance operated at room temperature, the dominant contribution is due to the coupling with a $^{13}$C nuclear spin bath with characteristic time $T_2^*\sim$~$\mu$s~\cite{Pham12}, while in isotopically purified diamond $T_2^*$ reaches $\sim100$~$\mu$s~\cite{Balasubramanian09}. Dephasing can be mitigated via dynamical decoupling, attaining coherence times typically limited by $T_2\sim0.5\,T_1$, with longitudinal relaxation times $T_1 \sim 6$~ms at room temperature and $T_1\sim 1$~s  at cryogenic temperature ($T=77$~K)~\cite{Bar-Gill13,Jarmola12}.
At room temperature, the triplet ground-state population ---distributed according to Maxwell-Boltzmann distribution--- can be transferred to the excited levels by irradiation of the center with green laser light ($532$~nm) through a process involving a combination of radiative absorption and non-radiative relaxation processes that also entails vibronic bands. A direct spin-preserving radiative decay from the excited to the ground level (with zero-phonon-line of $637$-nm wavelength) is accompanied by  
non-radiative non-spin-preserving decay through the long-lifetime singlet levels $^1A_1$ and $^1E$. 
The different decay rates for the different spin projections in the non-radiative decay channel yield the optical initialization of the system into the $m_S=0$ state of the ground level. The same decay mechanism enables spin state readout thanks to different photoluminescence intensities of the $m_S=0$ and $m_S=\pm 1$ states. A simplified sketch of a typical experimental setup for NV control is in Fig.~\ref{fig}(b).

\subsection{Magnetic field sensing}\label{magn}
Among sensing applications of NV centers, magnetometry---{\it i.e.}, the measurement of magnetic field strength and orientation---has received by far the most attention. As also for magnetometers based on gaseous alkali atoms, all the sensing protocols of NV magnetometers essentially reside within the measurement of the Zeeman splitting induced by an external magnetic field. The most basic scheme consists in the direct measurement of the transition frequency $|0\rangle\rightarrow|1\rangle$, via optically-detected electron spin resonance (ESR). This method has enabled the measurement of DC magnetic field, including {\it e.g.}  mapping of magnetic distributions with scanning probe 
magnetometers~\cite{Rondin13}, magnetic imaging in living cells with sub-cellular spatial resolution~\cite{LeSage13}, and noninvasive detection of action potentials with single-neuron sensitivity~\cite{Barry16}. This method is inherently limited to magnetic fields with amplitude and orientation such that the spin quantization axis remains aligned with the NV symmetry axis: since the NV readout relies on spin-dependent photoluminescence intensity, the spin mixing induced by orthogonal components of the magnetic field reduces the contrast of optically-detected ESR~\cite{Tetienne12}. However, the concomitant decrease of photoluminescence observed in the presence of any off-axis field has been used to perform all-optical magnetic field mapping~\cite{Tetienne12}. Vectorial magnetometry is possible with NV ensembles by exploiting the four different possible orientations of the NV centers in the diamond crystal~\cite{Maertz10,Pham11}.

For advanced sensing applications, a more precise determination of the transition frequency can be achieved with interferometric techniques. The basis of these techniques is Ramsey interference,~\cite{Ramsey50}, where the spin is prepared in a superposition state $|0\rangle+e^{i\varphi_0}|1\rangle$ and the spin phase $\varphi$ accumulated during the interrogation time reflects the addressed transition frequency, which depends on the external field to be measured. The Ramsey scheme is sensitive to static, slowly varying, or broadband near-DC signals~\cite{Degen17}. 
The sensitivity achieved with DC magnetometry schemes is limited by the dephasing time $T_2^*$. 

Higher sensitivity can be attained when measuring AC magnetic fields, by implementing a dynamical decoupling (DD) of the NV spin from its environment, thus prolonging the NV coherence time. The elementary DD protocol is Hanh's echo, where a $\pi$ pulse  applied halfway of the spin precession time reverses the spin evolution, so that the phase accumulated in the two time segments cancels out. This concept can be extended to multiple refocusing $\pi$ pulses, as shown in the next Section. The bandwidth of DD protocols usually extends up to $\sim 10$~MHz~\cite{Laraoui10,Steinert13,Schmitt17}, although interesting sensing schemes for signal detection up to $\sim100$~MHz~\cite{Aslam17} and even GHz~\cite{Joas17} have also been proposed. AC sensitivity is limited by the coherence time (also referred to as transverse relaxation time) $T_2$---due to homogeneous fluctuating fields, which can exceed $T_2^*$ by two orders of magnitude~\cite{DeLange10}.

\section{Magnetic spectroscopy}\label{sec:spectoscopy}

The resonant driving techniques developed for nuclear magnetic resonance (NMR)~\cite{Abragam61} in the first part of the previous century are still relevant for the development of novel techniques~\cite{Dobrovitski13}: In this section we explore some recent proposals that have taken NV magnetometry to new limits, focusing on pulsed DD protocols.

The spectral characterization of a magnetic field can be conducted through the analysis of relaxation and dephasing processes occurring to the probe system itself due to the interaction with the target field.
Relaxometry consists in the measurement of the relaxation rate $\Gamma$ of the sensor that is connected with the spectral density of a signal $S(\omega)$, linearly in the first order approximation. This method---introduced in NMR \cite{Kimmich04}, and also applied to superconductive qubits~\cite{Bialczak07,Lanting09,Bylander11,Yan13}---has been exploited with NV sensors to investigate especially high-frequency noise, through on the measurements of the $T_1$ relaxation time~\cite{Steinert13,Myers14,Rosskopf14,Vandersar15,Romach15,Stark17}. 

The alternative approach relies on the systematic analysis of the sensor decoherence under a set of DD control protocols~\cite{Viola98,Cywinski08,Faoro04,Almog11,Yuge11,Alvarez11,Young12}. 
Pulsed DD protocols, based on the Hahn's spin echo sequence, consist in sequences of $\pi$ pulses that repeatedly flip the qubit spin, hence reversing  its evolution, as sketched in Fig.~\ref{fig}(c): they realize narrow frequency filters that select only a specific coupling and frequency to be probed, while decoupling the sensor from the rest of the environment, hence extending the coherence time of the qubit to increase the measurement precision~\cite{Degen17}.

In several NV implementations, dephasing  
can be  modeled as due to a classical stochastic noise source~\cite{Reinhard12,HernandezGomez18}. In the presence of pure classical dephasing~\cite{Alvarez11,Biercuk11, Kotler11}, the DD control protocols can be effectively captured by the filter function approach~\cite{Degen17}. The control field can be described by a modulation function $y_n(t)$ with a sign switch at the position of each $\pi$ pulse, indicating the direction of time evolution, forward or backward.
The NV spin phase $\varphi$ acquired during the sensing time $T$ is mapped into
the residual population of the state $\ket{1}$:
\begin{equation}
P(T) = \frac{1}{2}\left( 1 + W(T)) \right),
\end{equation}
where $W$ is the qubit coherence. 
Since the source of dephasing is the nuclear spin bath that couples weakly to external magnetic target fields due the small nuclear magnetic moment, the qubit coherence $W$ can be factorized in two contributions, due to the external field to be measured $W^\mathrm{(ac)}$ and to the noise $W^\mathrm{(NSD)}$. 
In the presence of a target AC magnetic field $b(t)\equiv b\,f(t)$, the phase acquired under the action of the control field, $\varphi=\int_0^T\gamma\,b(t)y_n(t)\,dt$, modulates the qubit coherence  as $W^\mathrm{(ac)}(T) = \cos \varphi(T)$. 
On the other hand, the interaction with a detrimental noise source, with an associated noise spectral density~(NSD)~$S(\omega)$, tends to destroy the qubit coherence $W^\mathrm{(NSD)}(T) = e^{-\chi_n(T)}$, where the decoherence function
$\chi_n(T) = \int \dfrac{\mathrm{d}\omega}{\pi\omega^2} S(\omega) \left| Y_{n,T}(\omega) \right|^2$~\cite{Cywinski08,HernandezGomez18,Biercuk11}
is the convolution between the NSD and the filter function $Y_{n,T}(\omega) = 1 + (-1)^{n+1}e^{-i\omega T} + 2\sum_{j=1}^{n}(-1)^{j}e^{-i\omega \delta_j}$, where $\delta_j$ is the position of the $j$-th $\pi$ pulse\footnote{Note that $S(\omega)$ is expressed in MHz, and $Y_{n,T}(\omega)$ is dimensionless. An alternative formulation of the filter function is reported in~\cite{Degen17}, where $\chi_n(T) = \frac{1}{\pi}\int \mathrm{d}\omega S(\omega) \left| \mathcal{Y}_{n,T}(\omega) \right|^2$ with a filter function $\mathcal{Y}_{n,T}(\omega) \equiv \int_0^T \mathrm{d}t\, e^{-i\omega t} y_n(t)$ expressed in units of 1/Hz.} (see Fig.~\ref{fig}(d)).
For a single-spin sensor, sensitivity---that is, the minimum detectable signal per unit time, can be quantified as~\cite{Degen17,Poggiali18}
\begin{equation}
\eta=\frac{e^{\chi_n(T)}}{|\phi(T))|}\sqrt{T},
\end{equation} where $\phi=\varphi/b$ is the acquired phase per unit field.
Effective sensing thus relies on the identification of the optimal filter function that minimizes sensitivity, requiring the twofold tasks of capturing the target signal and rejecting unwanted noise, which may be even conflicting when signal and noise have mutual spectral content. Huge effort has been devoted in the last years to develop a suite of DD-based spin manipulation protocols,
opportunely tailored for mitigating the effect of different noise sources of decoherence, optimizing the decoupling performance, narrowing the spectral response and suppressing signal harmonics and sidebands, as well as  compensating pulse errors~\cite{Cywinski08}.

A large family of DD protocols is constituted by periodically-structured train of pulses. An example is the Carr-Purcell (CP) sequence
~\cite{Carr54}, formed by $n$ equidistant $\pi$ pulses, which enables the detection of monochromatic AC fields with periodicity commensurate with the interpulse delaytime $2t_1$: 
this pulse train acts as a narrow quasi-monochromatic and tunable filter, where $t_1$ selects the pass-band frequency, while $n$ determines the filter width~\cite{Taylor08,Maze08}.  As an extension of this scheme, XY-N sequences
~\cite{Gullion90} are designed to improve robustness against detuning and imperfections of the $\pi$-pulses, by symmetrically rotating their relative phase (see Fig.~\ref{fig}(c)). 
Periodic protocols have been demonstrated to be ideal for decoupling from environments with soft frequency-cutoff, such as P1 centers electronic spin bath~\cite{DeLange10,Wang12, Knowles14}. They have been exploited in NV-sensor settings to detect and characterize individual $^{13}$C nuclear spins~\cite{Zhao12,Taminiau12,Kolkowitz12l}, demonstrating the possibility of sensing single nuclear spins placed a few nanometers apart from the NV center (see Fig.~\ref{fig}(e)), and to detect and characterize proton spins of molecules in organic samples placed some nanometers outside the diamond~\cite{Staudacher13,Shi15}.

Further developments are introduced by non-equispaced and concatenated sequences. 
The Uhrig DD (UDD) protocol~\cite{Uhrig07}, composed by a set of $n$ $\pi$-pulses with interpulse delaytime 
 $\delta_j=\sin^2\left[ \pi j/(2n+2) \right]$ and $j\in\{1,\dots,n\}$,  has been successfully employed to detect $^{13}$C dimers~\cite{Zhao11} inside diamond, as it highly suppresses the effect of coupling to single nearby nuclei (see Fig.~\ref{fig}(f)). 
Nested sequences composed by repeated blocks of periodic or aperiodic $\pi$ pulses were designed to facilitate the discrimination of single nuclear spins~\cite{Zhao14}, thanks to high frequency selectivity and moderate peak strength of the filter function.  
Adding phase rotation within each block improves the robustness against pulse errors, an example of which being the adaptive XY-N (AXY-N) sequence~\cite{Casanova15}, where each block is a Knill pulse formed by $M=5$ equidistant $\pi$-pulses~\cite{Casanova15,Souza11KDD} (see Fig.~\ref{fig}(g)). The analytic expressions of the filter functions of the mentioned protocols are reported in Table~\ref{table1}.

\begin{table}[h!]
  \begin{center}
    \caption{Filter function $Y_{n,T}(\omega)$ of ESR, Ramsey free induction decay (FID), and selected pulsed DD protocols~\cite{Cywinski08, Degen17}. $n$, the total number of pulses; $T$, the sensing time; $t_\pi$, the pulse length; $c$, a normalization constant. For nested sequences, $N$ is the number of repetitions of a $M$-pulse block; the position in time of each pulse is $t_{m,n} =[(2n-1)/2+r_m](T/N)$ with $n=1,2,...,N$ and $m = 1,2,...,M$}
    \label{tab:table1}
    \renewcommand{\arraystretch}{1.4} 
    \begin{tabular}{lll}
      \hline\hline
      Protocol & $|Y_{n,T}(\omega)|^2$ & Reference\\
      \hline
      ESR &
      $
      \left\{
\begin{array}{rl}
1/c,&\mbox{for }2\pi/T<\omega <t_\pi \\
0,&\mbox{otherwise}
\end{array}
\right.
\label{table1}
$
     & \cite{Dreau11}\\
      FID & $2 \sin^2{(\frac{\omega T}{2})}$ & \cite{Degen17}\\
      CP (even $n$) & $8[\sec{(\frac{\omega T}{2n})}\sin(\frac{\omega T}{2})\sin^2{(\frac{\omega T}{4n})}]^2$ & \cite{Carr54}\\
      UDD & $\frac{1}{2}|\sum_{k=-n-1}^{n} (-1)^{k}\exp{[\frac{i\omega T}{2}\cos{(\frac{\pi k}{n+1})}]}|^2$& \cite{Uhrig07}\\ 
      Nested sequence$^\dagger$ & $16\frac{\sin^2{(\omega T/2)}}{\cos^2{(\omega T/2N)}}\left[\cos^2{(\omega T/4N)}-\cos{(r\omega T/N)}\right]^2$&\cite{Zhao14}     \\     
      \hline\hline
      \multicolumn{3}{l}{$^\dagger$ for $M=3$, $r_2=0$ and $r_3=-r_1\equiv r$ with $r\in[-\frac{1}{2},\frac{1}{2}]$}
    \end{tabular}
  \end{center}
\end{table}

The search for optimized DD protocols has also benefited from the application of Quantum Optimal Control (QOC) theory~\cite{Glaser15,Rembold20}, which exploits numerical optimization methods to find the best control field that opportunely steers the system dynamics towards a desired objective, subject to some control restrictions determined by physical and experimental constrains. In the very last years, QOC has shown a number of interesting results on NV settings~\cite{Haberle13,Scheuer14,Nobauer15,Poggiali18,Ziem19}. By introducing the Fisher information of the measurement as the cost function of the optimization~\cite{Poggiali18}, QOC can naturally take into account both the signal of interest and the environmental noise to find the optimal DD spin manipulation protocol (for example, defining the optimal  pulse distribution). QOC-DD schemes have been successfully employed in  single-qubit sensing  of complex AC fields, demonstrating a significant sensitivity improvement compared to the CP scheme~\cite{Poggiali18}.

\subsection{Noise spectroscopy}
A good knowledge of the environment is imperative in order to improve the sensor capabilities, either by allowing to strategically filter out the unwanted noise components~\cite{Poggiali17}, or by using part of the environment as ancillary systems~\cite{Goldstein11}.

In type-IIa diamond, the Carbon nuclear spin environment can be divided into a small set of strongly-coupled nuclei, and a large nuclear spin bath. The coherent coupling with the resolved nuclei can be characterized, as mentioned before, using periodic DD sequences~\cite{Zhao12,Taminiau12,Kolkowitz12l}. On the other hand, the interaction with the collective bath is responsible for the NV decoherence and its description is more involved. In the presence of strong bias magnetic field, the environment internal energy overcomes the typical NV-bath coupling strength. In this weak-coupling regime, the spin bath can be modeled as a classical stochastic field with NSD peaked at the $^{13}$C Larmor frequency~\cite{Alvarez11,HernandezGomez18}. 
However, the spectral characterization of noise can be quite challenging, and requires deconvolution analysis. 
The most common approach involves using CP-based sequences with large $n$, resulting in a filter function that can be approximated to a Dirac comb. Measuring a generalized coherence time~\cite{Yuge11,Alvarez11} allows the reconstruction of the NSD lineshape. 
When the noise is strong enough to destroy coherence in short times, using higher harmonics of the filter function for the NSD characterization will give cleaner and more accurate results~\cite{HernandezGomez18,10.1117/12.2531734}. 
For low bias fields  ($\leq 150$~G), the loss of NV coherence is due to the creation of entanglement between NV spin and the large environment: In this strong-coupling regime, the environment description in terms of classic noise is no longer valid~\cite{Reinhard12}, and the dynamics of the nuclear spin environment itself is affected by the control applied to the NV center electronic spin, due to the NV back action~\cite{HernandezGomez18}. 

In isotopically purified samples, paramagnetic impurities---especially P1 centers---dominate the NV dephasing. This electronic spin bath has been characterized with single-spin sensors and ensembles by combining either Hanh echo~\cite{Knowles14} or double quantum coherence magnetometry that employs the $m_s=\{+1,-1\}$ NV spin ground-state subspace~\cite{Bauch18}, with radiofrequency bath driving~\cite{DeLange12}.

\section{Summary and Prospect}

Diamond NV centers have been established as a prominent platform for a suite of quantum technology applications, among which quantum sensing is definitely the most mature. Quantum Control plays a crucial role in improving the sensor performance, by enhancing the sensor response to the target field to be measured, while protecting it against the remaining environment. Among QC strategies, the development of multi-pulse DD protocols for the NV spin manipulation has opened the way to impressive progress in magnetic spectroscopy, making possible the detection of ultra-thin magnetic fields such as that originated by single nuclei in the proximity of the NV sensor. 

DD techniques have been so far mainly employed for single isolated sensing qubits in bulk diamond. Major challenges and potential breakthroughs currently concern the application of DD spin manipulation protocols to two other relevant classes of NV magnetometers: scanning probe magnetometers --- which guarantee the best performance in terms of spatial resolution by virtue of the use of nano-fabricated diamond tips, and ensemble magnetometers --- providing enhanced signal-to-noise ratio thanks to the statistical averaging over multiple spins.
Albeit DD protocols have been demonstrated to be beneficial in multispin metrology~\cite{Pham12,Bar-Gill12} and in high-purity nanodiamonds~\cite{Knowles14}, often the poor spin coherence properties of NV-rich bulk diamond and nanodiamond NVs has  so far limited the application of these classes of settings to the measurement of strong DC fields based on optically-detected ESR or Ramsey interferometry, where sensitivity is limited by the dephasing time $T_2^*$, presently far away by orders of magnitude from the physical limit of $T_1/2$. This hindrance is not fundamental, and will be presumably overcome in the near future via improved synthesis techniques~\cite{EKIMOV2020161}, convenient experimental design ({\it e.g.}, magnetic gradient compensation and operation under strong bias magnetic fields to mitigate the effect of external electric-field gradients and  internal strain)~\cite{Barry20}, as well as increased collection efficiency~\cite{Aharonovich14}.
The application of DD spin manipulation protocols to scanning NV magnetometers and ensemble devices could pave the way  to reach fundamental metrology limits, and dramatically expand the range of envisioned nanoscale applications, for example enabling the detection of arbitrary individual spins in ensembles in the presence of environmental noise.

\section*{Conflict of Interest Statement}
The authors declare that the research was conducted in the absence of any commercial or financial relationships that could be construed as a potential conflict of interest.

\section*{Author Contributions}
SHG and NF wrote this review article and they are responsible for the content of the work.



\bibliographystyle{frontiersinHLTH&FPHY} 
\bibliography{../FiP-Biblio}


\section*{Figure captions}


\end{document}